\documentclass{iopconfser}
\usepackage{eurosym}
\usepackage{amssymb}
\usepackage{amsfonts,cite}
\usepackage{amsmath}
\usepackage{graphicx}
\usepackage{multirow}
\usepackage{xcolor}
\usepackage{epsfig}

\begin{document}

\title{Toda field theories and Calogero models associated to infinite Weyl groups}

\author{Andreas Fring}

\affil{Department of Mathematics, City St. George's, University of London, \\ Northampton Square, London EC1V 0HB, UK}

\email{a.fring@city.ac.uk}

\begin{abstract}
Many integrable theories can be formulated universally in terms of Lie algebraic root systems. Well-studied are conformally invariant scalar field theories of Toda type and their massive versions, which can be expressed in terms of simple roots of finite Lie and affine Kac-Moody algebras, respectively. Also, multi-particle systems of Calogero-Moser-Sutherland type, which require the entire root system in their formulation, are extensively studied. Here, we discuss recently proposed extensions of these models to similar systems based on hyperbolic and Lorentzian Kac-Moody algebras. We explore various properties of these models, including their integrability and their invariance with respect to infinite Weyl groups of affine, hyperbolic, and Lorentzian types.
\end{abstract}

\section{Introduction}

The mathematical concept of symmetries plays a crucial role in describing many physical phenomena. Famously, local gauge symmetries for three of the four known fundamental forces are represented in the Standard Model by finite-dimensional Lie algebras: the electromagnetic force by u(1), the weak nuclear force by su(2), and the strong nuclear force by su(3). To incorporate gravity, it is a natural conjecture to consider enlarging the symmetry algebras in some form. In string theory, one of the leading candidates to incorporate gravity, infinite-dimensional Kac-Moody algebras \cite{kacinfinite} are required; these algebras also appear in closely related conformal field theories \cite{goddard1986kac}. Progressing further, the symmetries in more modern versions of string theory and their extensions, such  type II theories and M-theory, are even larger and may be described by hyperbolic or Lorentzian Kac-Moody algebras \cite{west2000hidden,damour200210,englert2003s}.

Thus, in the pursuit of generalising the Standard Model, it seems very natural to consider theories with symmetries described by by increasingly general Lie algebras. In turn, subsectors of the Standard Model or models describing different types of physics may be formulated equivalently in terms of the roots associated with these algebras. Alternatively, one may also take the root space formulation as a starting point. The natural question we are asking here is whether versions of these theories based on finite or affine algebras can be extended to hyperbolic and Lorentzian algebras and what type of properties these models possess - in particular, whether they are integrable.

For instance, Toda field theories constitute a class of integrable systems that have been extensively studied and are well understood on the classical as well as on the quantum level. They are field theories consisting of $\ell$ scalar fields $\phi^i$, $i=1,\ldots,\ell$, expressed in terms of the root vectors $\alpha_i$  taken in an $\ell$-dimensional representation
\begin{equation}	
	{\cal L}_{\mathbf{g}}= \frac{1}{2}  \partial_\mu \phi  \partial^\mu \phi  - \frac{g}{\beta^2 }  \sum_{i=a}^{r}  n_i  e^{ \beta \alpha_i \cdot   \phi}  .\label{lagr}
\end{equation}	
Depending now on the choice of the lower limit $a$ in the sum and the defining set of the coupling constant $\beta $, one obtains theories with fundamentally different properties. By taking $a=1$ and $\beta \in \mathbb{R}$, so that the sum extends over the $r$ simple roots of a semi-simple Lie algebra $\bf{g}$, and defining the $n_i$ as the expansion coefficients of the highest root $\theta= \sum_{i=1}^r n_i \alpha_i$, the theory is conformally invariant and therefore massless. Instead, taking $a=0$, $\beta \in \mathbb{R}$ with $\alpha_i = - \theta$, $n_0=1$ produces an affine Toda field theory which is a massive theory that is extremely well understood up the level of an exact quantum scattering matrix known to all orders in perturbation theory \cite{BCDS,PD1,FO,q2}. While the scattering theory with $\beta \in \mathbb{R}$ does not involve any backscattering, taking instead $\beta \in i \mathbb{R}$ leads to nondiagonal S-matrices that allow for backscattering and complex soliton solutions on the classical level \cite{Holl}. Here we are addressing the question: what properties do the theories have when the sum in (\ref{lagr}) is extended to $a=-1,-2,-3,\ldots$?

\noindent Another well studied type of integrable theories, formulated in terms of generic Lie algebraic roots $\alpha \in \mathbb{R}^\ell $, are the multi-particle Calogero models
\begin{equation}
	H = \frac{1}{2} p^2 +  \sum_{  \alpha \in \Delta_{ \bf{g}   }} \frac{c_{\alpha}}{( \alpha \cdot q )^2 } = \frac{1}{2} p^2 +  \sum_{i=1}^r  \sum_{n=1}^{ h } \frac{ c_{in}}{\left[ \sigma^n ( \gamma_i) \cdot q \right]^2 } , \label{calodef}
	\end{equation} 
with coordinates $q=(q_1,\ldots, q_\ell)$  and momenta $p=(p_1,\ldots, p_\ell)$. In contrast to the Toda theories, in these models one requires not just the simple roots but the entire root space $\Delta_{ \bf{g}   }$. The coupling constants $c_{\alpha}$, $c_{in}$ are taken to be real. Here we present two equivalent representations, in which, in the second, we sum over $r$ orbits of the Coxeter element $\sigma$ of order $h$ acting on a representative $\gamma_i$, which equals either $\alpha_i$ or $-\alpha_i$. Similar integrable models such as Calogero or Calogero-Moser-Sutherland are obtained by replacing $x^2$ in the denominator by $\sin^2(x)$, $\sinh^2(x)$ or $\text{sn}^2(x) $, respectively. The generalisation of the first variant in (\ref{calodef}) to hyperbolic algebras was first suggested in \cite{olafdon}. In general, we address here the question: what properties do the theories have when we extend the sum in (\ref{calodef}) over an infinite set of roots?

 Let us now first assemble the necessary mathematical tools and subsequently discuss some physical properties of the generalised models.

\section{N-extended Lorentzian Kac-Moody algebras}
Central objects in the defining relations of the models are the $\ell$-dimensional root vectors, whose mutual inner products are encoded in the Cartan matrix $K_{ij}=2 \alpha _{i}\cdot \alpha _{j}/\alpha^2_{j}$. For semi-simple Lie algebras, $i,j=1,\ldots,r$, it is positive definite, and for affine Kac-Moody algebras, $i,j=0,1,\ldots,r$,  it is positive semi-definite. In both cases $ \left( \cdot \right) $ denotes the standard Euclidean inner product. Algebraically, the matrix occurs in the commutation relations of the associated Chevalley generators satisfying the Serre relations
\begin{equation}
	\lbrack
	H_{i},H_{j}]=0,~~[H_{i},E_{j}]=K_{ij}E_{j},~~[H_{i},F_{j}]=-K_{ij}F_{j},~~[E_{i},F_{j}]=\delta _{ij}H_{i},
\end{equation}
\begin{equation}
	\lbrack E_{i},\ldots ,[E_{i},E_{j}]\ldots ]=0, ~~\ \ \ [F_{i},\ldots
	,[F_{i},F_{j}]\ldots ]=0 .
\end{equation}
For the extended algebras we have to introduce a Lorentzian inner product defined here  for two $\ell +2m$ dimensional vectors $x=(x_{1},\ldots
,x_{\ell +2m})$ and $y=(y_{1},\ldots ,y_{\ell +2m})$ as
\begin{equation}
	x\cdot y:=\sum\limits_{i =1}^{\ell }x_{i }y_{i
	}-\sum\limits_{i =1}^{m}\left( x_{\ell +2i -1}y_{\ell +2i
	}+x_{\ell +2\i }y_{\ell +2i -1}\right) .  \label{vecmult}
\end{equation}
Following the idea of the proposal in \cite{gaberdiel2002class}, we then define here an extended root system as
	\begin{equation}
	\alpha ^{(n)}:=\left\{ \alpha _{1},\ldots ,\alpha _{r},\alpha
	_{0}=k_{1}-\theta ,\ldots
	,\alpha _{-j}=k_{j-1}-\left( k_{j}+\bar{k}_{j}\right) \right\} , \quad j=2,\ldots, n. \label{roots}
\end{equation}
For each $i$ the vectors $k_i$ and $\bar{k}_i$ represent a two-dimensional Lorentzian lattice. With $k_i=(1,0)$, $\bar{k}_i=(0,-1)$ and the inner product (\ref{vecmult}) with $\ell=0$ and $m=1$, the $k_i$, $\bar{k}_i$ turn out to be mutually orthogonal null vectors
\begin{equation}
	k_i \cdot k_i =   	\bar{k}_i \cdot \bar{k}_i  =0, \qquad  k_i \cdot \bar{k}_j  = \delta_{i,j} .
\end{equation}
With these definitions, we can compute the corresponding Cartan matrix, which can be encoded into a Dynkin diagram with the standard rules: associating to each simple root a node and joining up two nodes $i$ and $j$ with $\lambda$ links when $K_{ij}=-\lambda$. Similarly to the classification scheme of semi-simple and affine Lie algebras \cite{kacinfinite}, Dynkin diagrams can also be used for the classification, or at least the definition, of extended systems. For the finite-dimensional and affine algebras, we use the notation ${\bf g}$ and ${\bf g}_0$, respectively, and for the n-extended versions ${\bf g}^{(n)}$. The following general definitions have been proposed:

Def.: A {\em  hyperbolic Kac-Moody algebra} (\underline{Lorentzian Kac-Moody algebra}) is defined as an algebra such that the deletion of {\em any one node} (\underline{at least one node}) from its Dynkin diagram results in a possibly disconnected set of connected Dynkin diagrams, each of which is of finite type, except for at most one that is of affine type. Popular examples in string theory are for instance the so-called extended $\mathbf{E}_{8}^{(1)} $-algebra, or simply $\mathbf{E}_{10}$, and for the latter the so-called over-extended $\mathbf{E}_{8}^{(2)} $-algebra, or  simply $\mathbf{E}_{11}$. We easily convince ourselves that these definitions hold for the corresponding Dynkin diagrams:

	\begin{center}	
	\setlength{\unitlength}{0.45cm} 
	\begin{picture}(16.,2.5)(3.8,4)
		\thicklines
		\put(0.5,5.0){\Large{$\mathbf{E}_{8}^{(1)} \equiv \mathbf{E}_{10}:$}}
		\put(7.0,5){\Large{$\bullet$}}
		\put(7.4,5.2){\line(1,0){0.9}}
		\put(6.8,4.5){{$\small{\alpha_{1}}$}}
		\put(8.0,5){\Large{$\bullet$}}
		\put(8.4,5.2){\line(1,0){0.9}}
		\put(8.0,4.5){{$\small{\alpha_{3}}$}}
		\put(9.0,5){\Large{$\bullet$}}
		\put(9.4,5.2){\line(1,0){0.9}}
		\put(9.0,4.5){{$\small{\alpha_{4}}$}}
		\put(9.0,6){\Large{$\bullet$}}
		\put(9.2,5.2){\line(0,1){0.9}}
		\put(9.5,6.2){{$\small{\alpha_{2}}$}}
		\put(10.0,5){\Large{$\bullet$}}
		\put(10.4,5.2){\line(1,0){0.9}}
		\put(10.0,4.5){{$\small{\alpha_{5}}$}}
		\put(11.0,5){\Large{$\bullet$}}
		\put(11.4,5.2){\line(1,0){0.9}}
		\put(11.0,4.5){{$\small{\alpha_{6}}$}}
		\put(12.0,5){\Large{$\bullet$}}
		\put(12.4,5.2){\line(1,0){0.9}}
		\put(12.0,4.5){{$\small{\alpha_{7}}$}}
		\put(13.0,5){\Large{$\bullet$}}
		\put(13.2,5.2){\line(1,0){0.9}}
		\put(12.8,4.5){{$\alpha_{8}$}}
		\put(14.0,5){\Large{$\bullet$}}
		\put(14.4,5.2){\color{black}{\line(1,0){0.9}}}
		\put(13.8,4.5){{$\small{\alpha_{0}}$}}
		\put(15.0,5){\Large{$\color{black}{ \bullet}  $}}
		\put(14.8,4.5){{$\small{ \color{black}{\alpha_{-1}}}$}}
	\end{picture}	
	\begin{picture}(11.00,2.5)(3.5,4)
		\thicklines
		\put(0.5,5.0){\Large{$\mathbf{E}_{8}^{(2)} \equiv \mathbf{E}_{11}:$}}
		\put(7.0,5){\Large{$\bullet$}}
		\put(7.4,5.2){\line(1,0){0.9}}
		\put(6.8,4.5){{$\small{\alpha_{1}}$}}
		\put(8.0,5){\Large{$\bullet$}}
		\put(8.4,5.2){\line(1,0){0.9}}
		\put(8.0,4.5){{$\small{\alpha_{3}}$}}
		\put(9.0,5){\Large{$\bullet$}}
		\put(9.4,5.2){\line(1,0){0.9}}
		\put(9.0,4.5){{$\small{\alpha_{4}}$}}
		\put(9.0,6){\Large{$\bullet$}}
		\put(9.2,5.2){\line(0,1){0.9}}
		\put(9.5,6.2){{$\small{\alpha_{2}}$}}
		\put(10.0,5){\Large{$\bullet$}}
		\put(10.4,5.2){\line(1,0){0.9}}
		\put(10.0,4.5){{$\small{\alpha_{5}}$}}
		\put(11.0,5){\Large{$\bullet$}}
		\put(11.4,5.2){\line(1,0){0.9}}
		\put(11.0,4.5){{$\small{\alpha_{6}}$}}
		\put(12.0,5){\Large{$\bullet$}}
		\put(12.4,5.2){\line(1,0){0.9}}
		\put(12.0,4.5){{$\small{\alpha_{7}}$}}
		\put(13.0,5){\Large{$\bullet$}}
		\put(13.2,5.2){\line(1,0){0.9}}
		\put(12.8,4.5){{$\alpha_{8}$}}
		\put(14.0,5){\Large{$\bullet$}}
		\put(14.4,5.2){\color{black}{\line(1,0){0.9}}}
		\put(13.8,4.5){{$\small{\alpha_{0}}$}}
		\put(15.0,5){\Large{$\color{black}{ \bullet}  $}}
		\put(14.8,4.5){{$\small{ \color{black}{\alpha_{-1}}}$}}		
		\put(15.2,5.2){\color{black}{\line(1,0){0.9}}}
		\put(16.0,5){\Large{$\color{black}{ \bullet}  $}}
		\put(15.9,4.5){{$\small{ \color{black}{\alpha_{-2}}}$}}
	\end{picture}	
\end{center}

 From the $\mathbf{E}_{10}$-diagram, we may remove any node and the definition holds. From the $\mathbf{E}_{11}$-diagram, one may removed many of the nodes so that the remaining disconnected diagrams are of finite or of affine type. However, it does not hold for the last node associated with $\alpha_{-2}$, so that $\mathbf{E}_{11}$ is not hyperbolic according to the above definition. Also note that our definitions for the roots is a special case that can easily be generalised. The simple construction principle here is to add $k_1$ in the highest root, which does not change  anything for the affine part of the diagram, as it is a null vector, but it can be used ``as a hook" to connect to the next root that contains $\bar{k}_1$. This linking procedure can be used for any other node in order to construct all kinds of extended Dynkin diagrams. Here we use it to keep attaching roots to the new roots.  

As for the finite and affine case, we also need the weight lattice that consists of the fundamental weight vectors $\lambda _{i}^{(n)}$ orthogonal to the root vectors  $\lambda _{i}^{(n)}\cdot \alpha _{j}^{(n)}=\delta _{ij}$ for $
i,j=-n,\ldots,0,1,\ldots ,r$ such that $\lambda _{i}^{(n)}=\sum_{j=-n}^{r}K_{ij}^{-1}\alpha _{j}^{(n)}
$, $K_{ij}^{-1}=\lambda _{i}^{(n)}\cdot \lambda _{j}^{(n)}$. In \cite{fring2019n} we constructed the closed formulae
\begin{eqnarray}
	\lambda _{i}^{(n)} &=&\lambda _{i}^{f}+n_{i}\lambda _{0}^{(n)},~~\ \
	i=1,\ldots ,r,  \label{l1} \\
	\lambda _{0}^{(n)} &=&\bar{k}_{1}-k_{1}+\frac{1}{n}\sum\limits_{i=2}^{n}%
	\left[ k_{i}+(n+1-i)\bar{k}_{i}\right] ,  \label{l0} \\
	\lambda _{-j }^{(n)} &=& \!\!\!\frac{(1-j )}{n}\sum\limits_{i=2}^{n}\left[ k_{i}+(1-i)\bar{k}_{i}%
	\right] + \! \sum\limits_{i=2}^{j -1}\left[ k_{i}+(1-i)\bar{k}_{i}\right]
	+(1-j )\sum\limits_{i=j }^{n}\bar{k}_{i}, \quad 	j=1,\ldots ,n.
\end{eqnarray}
Having obtained generic expressions for the weight vectors, we can calculate the $n$-extended Weyl vector by summing over all weights
\begin{eqnarray}
	\rho ^{(n)} &=&\sum_{j=-n}^{r}\lambda _{j}^{(n)} , \label{rn} \\
	&=&\rho ^{f}+h\bar{k}_{1}-(1+h)k_{1}+\sum\limits_{i=2}^{n}\left[ \left( 
	\frac{h}{n}+\frac{n+1-2i}{2}\right) k_{i}  +\frac{(n+1-i)(2h+n(1-i))}{2n}\bar{k%
	}_{i}\right] .
\end{eqnarray}
The Lorentzian inner product of this vector with itself yields an n-extended version of the so-called Freudenthal-de Vries strange formula
\begin{equation}
	\rho ^{(n)}\cdot \rho ^{(n)}=\frac{h(h+1)r+n(n^{2}-1)}{12}-\frac{h(h+n)(1+n)%
	}{n} ,   \label{FdVsf}
\end{equation}
which for $n=-1$ reduces to the well-known formula for the finite case
\begin{equation}
	\left( \rho ^{f}\right) ^{2}=\frac{h}{12}\text{dim}\mathbf{g=}\frac{h(h+1)r}{%
		12}.
\end{equation} 
For finite and affine Toda theories, the use of the $SO(3)$ principal subalgebra has turned out to be extremely instrumental in unravelling their properties. Its three generators, $J_\pm$, $J_3$, satisfying $\left[ J_{+},J_{-}\right] =J_{3}$, $\left[ J_{3},J_{\pm }\right]
=\pm J_{\pm }$, can be composed from the $r(h+1)$ generators $H_i$, $E_i$, $F_i$ in the Chevalley basis as
\begin{equation}
	J_{3}=\sum_{i=1}^{r}D_{i}H_{i},\quad J_{+}=\sum_{i=1}^{r}\! \sqrt{D_{i}}E_{i}, \quad
	J_{-}=\sum_{i=1}^{r} \! \sqrt{D_{i}}F_{i},\qquad  {D_{i}:=\sum_{j=1}^{r}K_{ji}^{-1}}.
\end{equation}	
For the extended algebras the same role is played by the $SO(2,1)$-principal subalgebra \cite{nicolai2001dio}, whose three generators $\hat{J}_\pm$, $\hat{J}_3$ satisfy $\left[ \hat{J}_{+},\hat{J}_{-}\right] =-\hat{J}_{3}$, $\left[ 
\hat{J}_{3},\hat{J}_{\pm }\right] =\pm \hat{J}_{\pm }$, and can be expressed as
\begin{equation}
	\hat{J}_{3}=-\sum_{i=1}^{r}\hat{D}_{i}H_{i},\quad \hat{J}_{+}=%
	\sum_{i=1}^{r}p_{i}E_{i},\quad \hat{J}_{-}=\sum_{i=1}^{r}q_{i}F_{i},\quad 
	 {\hat{D}_{i}:=\sum_{j=1}^{r}K_{ji}^{-1}} ,
\end{equation}
with $p_{i}q_{i}=\left\vert p_{i}\right\vert ^{2}=-\hat{D}_{i}$.  This immediately implies that
 \emph{%
	a necessary and sufficient condition for the existence of a }$SO(3)$\emph{-principal subalgebra or a }$SO(1,2)$\emph{-principal subalgebra is }$
D_{i}>0 $\emph{\ or }$\hat{D}_{i}<0$\emph{\ for all }$i$\emph{, respectively}. Given the inverse of the Cartan matrix, one may verify this condition by direct calculation in case-by-case computations. Alternatively, one can also make use of a {\emph necessary condition for the existence of a $SO(1,2)$ principle subalgebra is $\rho^2 <0$.}  This property is easily checked on a case-by-case basis using the Freudenthal de Vries strange formula (\ref{FdVsf}).
These properties can be used to compute the decomposition of an extended $\mathbf{g}^{(n)}$ algebra. Whenever $D_{k}= 0 $ in any $\mathbf{g}^{(n)}$ Dynkin diagram, the decomposition is obtained by removing the$k$-th node together with its attached links. For instance, for the doubly extended $D_{25}^{(2)}$, we obtain the reduced diagram $D_{25}^{(2)}=E_{7}^{(1)}\diamond \color{blue}{L}\diamond \color{red}{ D_{18}}$: 

\setlength{\unitlength}{0.48cm} 
\begin{picture}(14.00,6.0)(-5.5,3.3)
\thicklines

\put(-2.0,8.0){\Large{$\bullet$}}

\put(-1.5,8.1){{$\gamma_{7}$}}

\put(-1.0,7.0){\Large{$\bullet$}}
\put(-0.8,7.2){\line(-1,1){1.0}}
\put(-0.5,7.1){{$\gamma_{6}$}}

\put(0.0,6.0){\Large{$\bullet$}}
\put(0.2,6.2){\line(-1,1){1.0}}
\put(0.5,6.1){{$\gamma_{5}$}}

\put(0.0,4.0){\Large{$\bullet$}}
\put(-0.8,4.0){{$\gamma_1$}}

\put(1.2,5.2){\line(-1,-1){1.0}}
\put(1.2,5.2){\line(-1,1){1.0}}
\put(1.0,5){\Large{$\bullet$}}
\put(1.2,5.2){\line(1,0){1.0}}
\put(1.0,4.5){{$\gamma_2$}}

\put(2.0,5){\Large{$\bullet$}}
\put(2.2,5.2){\line(1,0){1.0}}
\put(2.0,4.5){{$\gamma_3$}}

\put(3.0,5){\Large{$\bullet$}}
\put(3.2,5.2){\line(1,0){1.0}}
\put(3.0,4.5){{$\gamma_4$}}

\put(4.0,5){\Large{$\bullet$}}
\put(4.2,5.2){\line(1,0){0.9}}
\put(4.0,4.5){{$\gamma_0$}}

\put(5.0,5){\Large{$\bullet$}}
\put(5.0,4.5){{$\gamma_{-1}$}}

\color{blue}{
	\put(6.0,5){\Large{$\bullet$}}
	\put(6.0,5.5){{$-\ell$}}}

\color{red}{
	\put(7.0,5){\Large{$\bullet$}}
	\put(7.2,5.2){\line(1,0){1.0}}
	\put(7.0,4.4){{$\beta_1$}}

	\put(8.0,5){\Large{$\bullet$}}
	\put(8.2,5.2){\line(1,0){1.0}}
	\put(8.0,4.4){{$\beta_2$}}

	\put(9.0,5){\Large{$\bullet$}}
	\put(9.2,5.2){\line(1,0){1.0}}
	\put(9.0,4.4){{$\beta_{3}$}}

	\put(10.0,5){\Large{$\bullet$}}
	\put(10.2,5.2){\line(1,0){1.0}}
	\put(10.0,4.4){{$\beta_{4}$}}

	\put(11.0,5){\Large{$\bullet$}}
	\put(11.2,5.2){\line(1,0){1.0}}
	\put(11.0,4.4){{$\beta_{5}$}}
	
	\put(12.0,5){\Large{$\bullet$}}
	\put(12.2,5.2){\line(1,0){1.0}}
	\put(12.0,4.4){{$\beta_{6}$}}
	
	\put(13.0,5){\Large{$\bullet$}}
	\put(13.2,5.2){\line(1,0){1.0}}
	\put(13.0,4.4){{$\beta_{7}$}}
	
	\put(14.0,5){\Large{$\bullet$}}
	\put(14.2,5.2){\line(1,0){1.0}}
	\put(14.0,4.4){{$\beta_{8}$}}
	
	\put(15.0,5){\Large{$\bullet$}}
	\put(15.2,5.2){\line(1,0){1.0}}
	\put(15.0,4.4){{$\beta_{9}$}}
	
	\put(16.0,5){\Large{$\bullet$}}
	\put(16.2,5.2){\line(1,0){1.0}}
	\put(16.0,4.4){{$\beta_{10}$}}
	
	\put(17.0,5){\Large{$\bullet$}}
	\put(17.2,5.2){\line(1,0){1.0}}
	\put(17.0,4.4){{$\beta_{11}$}}
	
	\put(18.0,5){\Large{$\bullet$}}
	\put(18.2,5.2){\line(1,0){1.0}}
	\put(18.0,4.4){{$\beta_{12}$}}
	
	\put(19.0,5){\Large{$\bullet$}}
	\put(19.2,5.2){\line(1,0){1.0}}
	\put(19.0,4.4){{$\beta_{13}$}}
	
	\put(20.0,5){\Large{$\bullet$}}
	\put(20.2,5.2){\line(1,0){1.0}}
	\put(20.0,4.4){{$\beta_{14}$}}
	
	\put(21.0,5){\Large{$\bullet$}}
	\put(21.2,5.2){\line(1,0){1.0}}
	\put(21.0,4.4){{$\beta_{15}$}}
	
	\put(22.0,5){\Large{$\bullet$}}
	\put(22.2,5.2){\line(1,1){1.0}}
	\put(22.2,5.2){\line(1,-1){1.0}}
	\put(22.6,5.1){{$\beta_{16}$}}
	
	\put(23.0,6.0){\Large{$\bullet$}}
	\put(22.7,6.6){{$\beta_{17}$}}
	
	\put(23.0,4.0){\Large{$\bullet$}}
	\put(22.7,3.4){{$\beta_{18}$}}}

\end{picture} 
\medskip

\noindent For the $k$-th node, marked with $- \ell$, we obtain $D_k =0$, which we indicate as an algebra by $L$. Another example with three $D_k =0 $ is for instance $A_{24}^{(10)} = E_{6}^{(3)}\diamond L\diamond A_{4}\diamond L^{2}\diamond A_{18}$.

The above mathematical toolkit suffices to discuss the properties of extended Toda theories, but in order to formulate theories that require the entire root space we also need to make use of the Weyl reflections associated to each of the $r+n +1$ extended roots $\alpha_i^{(n)}$ 
\begin{equation}
	\sigma_i(x) := x- (\alpha_i^{(n)} \cdot x ) \alpha_i^{(n)}, \qquad i=-n,\ldots,r , \label{Weyldef}
\end{equation} 
and the Coxeter element constructed from them
\begin{equation}
	\sigma^{(n)} :=\prod_{i=-n}^r \sigma_i .
\end{equation} 
Since Weyl reflections do not commute in general, the order in which the products are taken needs to be specified. The entire root space may then be constructed by requiring it to be invariant under all simple Weyl reflections, or alternatively it may be obtained from orbits of the associated Coxeter element acting upon the appropriate representatives. The order of the Coxeter elements, i.e.~the Coxeter number $h$, is naturally only finite for finite dimensional algebras. Next, we build models from these root systems.

\section{$\mathcal{L}_{_{\mathbf{\mathring{g}}_{-n}}}$-extended Lorentzian Toda field theory}

Next, we introduce new models based on perturbing the Lagrangian  $\mathcal{L}_{_{\mathbf{g}}}$ as defined in (\ref{lagr}) for $a=1$. By consecutively adding specific roots, we obtain massless and massive models in an alternating fashion. The first examples follow the construction procedure 
\[
\mathcal{L}_{_{\mathbf{g}}}\overset{\alpha _{0}}{\rightarrow }\mathcal{L}%
_{_{\mathbf{g}_{0}}}\overset{\alpha _{-1}}{\rightarrow }\mathcal{L}_{_{%
		\mathbf{g}_{-1}}}\overset{\alpha _{-2}}{\rightarrow }\mathcal{L}_{_{\mathbf{%
			\mathring{g}}_{-2}}}
\]
where $\mathcal{L}_{_{\mathbf{g}}}$ is a conformal Toda field theory. When adding a modified affine root $\alpha
_{0}=k-\sum\nolimits_{i=1}^{r}n_{i}\alpha _{i} $ we obtain $\mathcal{L}_{_{\mathbf{g}_{0}}} $, which is an affine Toda field theory. Adding next the Lorentzian root $\alpha _{-1}=-\left( k+\bar{k}\right) $, we obtain a conformal extension of affine Toda theory. Next, we add the Lorentzian root $\alpha _{-2}=\bar{k} $, obtaining
$\mathcal{L}_{_{\mathbf{\mathring{g}}_{-2}}} $ which is a massive Lorentzian Toda  
field theory. We have slightly modified the notation from ${\mathbf{g}}$ to ${\mathbf{\mathring{g}}}$, as we did not add a root specified in the set reported in (\ref{roots}).  Continuing with
\[
\begin{array}{ll}
	\alpha _{i}\equiv \text{simple roots of }\mathbf{g} & \text{for }j=1,\ldots
	,r \\ 
	\alpha _{-(2i-2)}=k_{i}-\sum\nolimits_{j=-(2i-3)}^{r}n_{j}\alpha _{j}~~~\ \
	\  & \text{for }i=1,\ldots ,n \\ 
	\alpha _{-(2i-1)}=-(k_{i}+\bar{k}_{i}) & \text{for }i=1,\ldots ,n
\end{array}%
\]%
we obtain a series of massless models related to the algebras 

\setlength{\unitlength}{0.58cm} 
\begin{picture}(14.00,3.5)(-1.0,7.5)
	\thicklines
	\put(0.5,9.0){\Large{$\mathbf{\mathring{g}}_{1-2n}:$}}
	\put(3.4,9.2){\Large{$\ldots$}}
	\put(4.5,9.2){\line(1,0){0.6}}
	\qbezier[18](5.3,9.4)(6.2,10.5)(7.2,9.4)
	\qbezier[18](7.3,9.4)(8.2,10.5)(9.2,9.4)
	\qbezier[18](12.3,9.4)(13.2,10.5)(14.2,9.4)
	\put(5.0,9.0){\Large{$\bullet$}}
	\put(5.4,9.2){\line(1,0){0.9}}
	\put(4.9,8.5){{$\alpha_0$}}
	\put(6.0,9){\Large{$\bullet$}}
	\put(6.2,9.2){\line(1,0){0.9}}
	\put(5.8,8.5){{$\alpha_{-1}$}}
	\put(7.0,9){\Large{$\circ$}}
	\put(7.4,9.2){\line(1,0){0.9}}
	\put(6.9,8.5){{$\alpha_{-2}$}}
	\put(8.0,9){\Large{$\bullet$}}
	\put(8.2,9.2){\line(1,0){0.9}}
	\put(8.0,8.5){{$\alpha_{-3}$}}
	\put(9.0,9){\Large{$\circ$}}
	\put(9.4,9.2){\line(1,0){0.7}}
	\put(9.2,8.5){{$\alpha_{-4}$}}
	\put(10.2,9.2){\Large{$\ldots$}}
	\put(11.4,9.2){\line(1,0){0.7}}
	\put(12.0,9){\Large{$\circ$}}
	\put(12.4,9.2){\line(1,0){0.9}}
	\put(11.6,8.5){{$\small{\alpha_{4-2n}}$}}
	\put(13.0,9){\Large{$\bullet$}}
	\put(13.2,9.2){\line(1,0){0.9}}
	\put(14.0,9){\Large{$\circ$}}
	\put(14.4,9.2){\line(1,0){0.9}}
	\put(13.6,8.5){{$\small{\alpha_{2-2n}}$}}
	\put(15.0,9){\Large{$\bullet$}}
\end{picture}      
                               
\noindent For each $n$, they can be made massive by adding the root
$
\alpha _{-(2n)}=-\sum\nolimits_{j=-(2n-1)}^{r}n_{j}\alpha _{j}
$,
leading to the Dynkin diagram

\setlength{\unitlength}{0.58cm} 
\begin{picture}(14.00,3.5)(-1.0,3.5)
	\thicklines
	\put(0.5,5.0){\Large{$\mathbf{\mathring{g}}_{-2n}:$}}
	\put(3.4,5.2){\Large{$\ldots$}}
	\put(4.5,5.2){\line(1,0){0.6}}
	\qbezier[18](5.3,5.4)(6.2,6.5)(7.2,5.4)
	\qbezier[18](7.3,5.4)(8.2,6.5)(9.2,5.4)
	\qbezier[18](12.3,5.4)(13.2,6.5)(14.2,5.4)
	\qbezier[18](14.3,5.4)(15.2,6.5)(16.2,5.4)
	\put(5.0,5){\Large{$\bullet$}}
	\put(5.4,5.2){\line(1,0){0.9}}
	\put(4.8,4.5){{$\alpha_0$}}
	\put(6.0,5){\Large{$\bullet$}}
	\put(6.2,5.2){\line(1,0){0.9}}
	\put(5.8,4.5){{$\alpha_{-1}$}}
	\put(7.0,5){\Large{$\circ$}}
	\put(7.4,5.2){\line(1,0){0.9}}
	\put(6.9,4.5){{$\alpha_{-2}$}}
	\put(8.0,5){\Large{$\bullet$}}
	\put(8.2,5.2){\line(1,0){0.9}}
	\put(8.0,4.5){{$\alpha_{-3}$}}
	\put(9.0,5){\Large{$\circ$}}
	\put(9.4,5.2){\line(1,0){0.7}}
	\put(9.2,4.5){{$\alpha_{-4}$}}
	\put(10.2,5.2){\Large{$\ldots$}}
	\put(11.4,5.2){\line(1,0){0.7}}
	\put(12.0,5){\Large{$\circ$}}
	\put(12.4,5.2){\line(1,0){0.9}}
	\put(11.6,4.5){{$\small{\alpha_{4-2n}}$}}
	\put(13.0,5){\Large{$\bullet$}}
	\put(13.2,5.2){\line(1,0){0.9}}
	\put(14.0,5){\Large{$\circ$}}
	\put(14.4,5.2){\line(1,0){0.9}}
	\put(13.6,4.5){{$\small{\alpha_{2-2n}}$}}
	\put(15.0,5){\Large{$\bullet$}}
	\put(15.2,5.2){\line(1,0){0.9}}
	\put(16.0,5){\Large{$\circ$}}
	\put(15.6,4.5){{$\small{\alpha_{-2n}}$}}
\end{picture}		

We have extended here the standard set of rules for Dynkin diagrams. Besides the solid links that label off diagonal Cartan matrix entries with $-1$, we have also introduced dotted links indicating entries of $1$, which is familiar when describing Dynkin diagrams for superalgebras. In addition, we have also included circles indicating zero length roots besides the bullets that label roots of length 2. 

One of the interesting properties to investigate is the mass spectrum of the $\mathbf{\mathring{g}}_{-2n}$-series, corresponding classically, as usual, to the coefficients of quadratic term of the scalar fields in the Lagrangian. For the series based on
$\mathbf{g} = E_8$, our results are depicted in the left panel of the following figure  

\begin{center}
	\includegraphics[width=7.7cm]{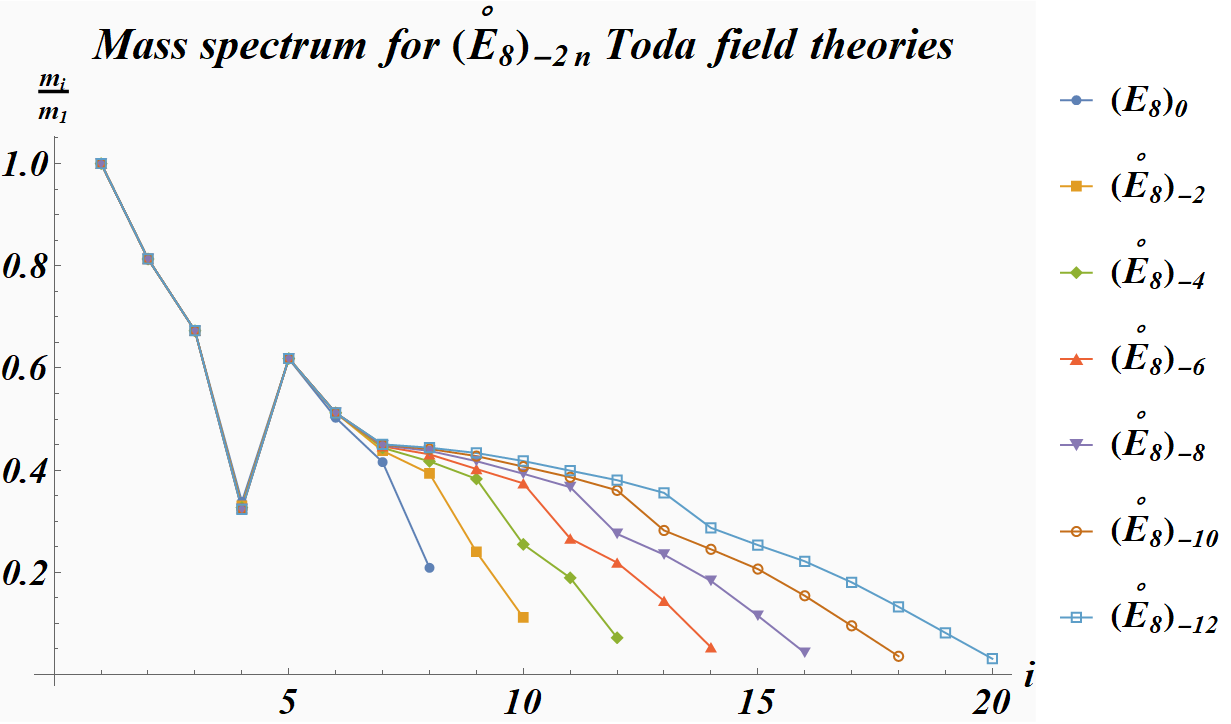}
	\includegraphics[width=7.7cm]{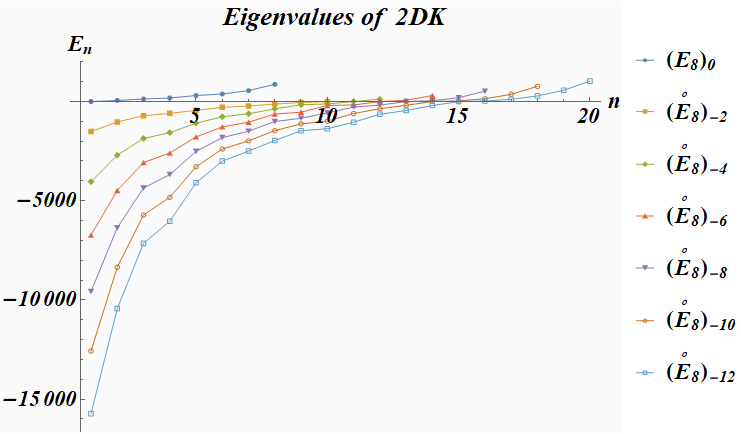}
\end{center}	
Notice the almost stable noncrystallographic $H_{4}$ compound for the first four masses. Of course we do not know at this stage whether the classical mass ratios are preserved after renormalization, as in the case of simply laced affine Toda theories.  The next important question to answer is whether the models are integrable or not. As discussed in more detail in \cite{fring2021lorentzian}, for a Lorentzian Toda field theory to be integrable all eigenvalues of the matrix $ P= 2 DK $
must be positive integers. These eigenvalues are depicted in the right panel of the previous figure. We observe that,  apart from the affine $E_8$ Toda theory, none of the further extensions to Lorentzian Toda theories are, in fact, integrable.

\section{Extended Calogero models}
As already mentioned, the Calogero models in (\ref{calodef}) require the knowledge of all roots. We present this extension now for the example of the doubly extended $A_2$-affine Kac Moody algebra with Dynkin diagram and Cartan matrix

\setlength{\unitlength}{0.58cm} 
\begin{picture}(13.00,5.)(0.0,2.5)
	\thicklines
	\put(1.5,5.0){\Large{ $    \mathbf{A}_2^{(2)}:$ }}
	\put(5.3,6){{$\small{\alpha_{1}}$}}
	\put(5.3,4){{$\small{\alpha_{2}}$}}
	\put(6.2,4.2){\line(0,1){2}}
	\put(6.0,6){\Large{$\bullet$}}
	\put(6.0,4){\Large{$\bullet$}}
	\put(7.2,5.2){\line(-1,1){0.9}}
	\put(7.2,5.2){\line(-1,-1){0.9}}
	\put(7.0,5){\Large{$\bullet$}}
	\put(7.4,5.2){\line(1,0){0.9}}
	\put(6.8,4.5){{$\small{\alpha_{0}}$}}
	\put(8.0,5){\Large{$\bullet$}}
	\put(8.2,5.2){\line(1,0){0.9}}
	\put(7.8,4.5){{$\small{\alpha_{-1}}$}}
	\put(9.0,5){\Large{$\bullet$}}
	\put(9.0,4.5){{$\small{\alpha_{-2}}$}}
	\put(11.0,5){$\!\!\! K_{ij}=2 \frac{\alpha_i \cdot \alpha_j}{\alpha_j \cdot \alpha_j}=\left(
		\begin{array}{ccccc}
			2 & -1 & 0 & 0 & 0 \\
			-1 & 2 & -1 & 0 & 0 \\
			0 & -1 & 2 & -1 & -1 \\
			0 & 0 & -1 & 2 & -1 \\
			0 & 0 & -1 & -1 & 2 \\
		\end{array}
		\right)_{ij}.$}
\end{picture}\\ 
A convenient representation of the roots is in a seven dimensional space
\begin{eqnarray}
		\alpha _{1} \!\!\! \!&=&\!\!\!\! \left( 1,-1,0;0,0,0,0 \right)
		,\,\, \alpha _{2}=\left(0,1,-1;0,0,0,0 \right) , \\
		\alpha _{0} \!\!\!\! &=& \! \!\!\! \left(-1, 0, 1; 1, 0, 0, 0 \right) , \,\,
		\alpha _{-1} =\left( 0,0,0;-1,1,0,0 \right) ,\,\,\alpha _{-2}=\left(
		0,0,0;1,0,-1,1\right) ,
\end{eqnarray}
when using the inner product (\ref{vecmult}) with $\ell =3$ and $m=2$. Demanding the length of an arbitrary root 
\begin{equation}
	\alpha = p \alpha_{-2} + q \alpha_{-1} + l \alpha_0 + m \alpha_1 + n \alpha_2 \in \Delta_{\mathbf{(A}_{2}\mathbf{)}_{-2}} \label{alphagen}, \qquad   p,q,q,m,n \in \mathbb{Z},
\end{equation}
to be 2, leads to the Diophantine equation
\begin{equation}
	l^2-l (m+n+q)+m^2-m n+n^2+p^2-p q+q^2 = 1 \qquad \Leftrightarrow \quad \alpha \cdot \alpha =2 . \label{Dioph}
\end{equation} 
The Weyl reflections defined in (\ref{Weyldef}) then generate the following symmetries of this equation
\begin{eqnarray}
	\!\!\!\!	&&  \!\!\!\!\sigma_{-2}(\alpha) : p \rightarrow q-p , \quad
	\sigma_{-1}(\alpha) : q \rightarrow l+p-q , \quad
	\sigma_{0}(\alpha) : l \rightarrow q+m+n-l ,  \label{LWeylref} \\
	\!\!\!\!	&&\!\! \!\! \sigma_{1}(\alpha) : m \rightarrow l+n-m , \quad 
	\sigma_{2}(\alpha) : n \rightarrow l+m-n .
\end{eqnarray} 
The left version of the potential (\ref{calodef}) is then by construction invariant under all of these reflections and may be cast into the form
\begin{equation}
	V(q) =   \!\!\!\!\!\!\! \!\!\!\!\!\!\!   \sum_{\begin{array}{cc}
			{ \small p,q,l,m,n=0 }\\
			{\small{\text{Diophantine equation}}}
	\end{array} }^\infty   \!\!\!\!   \frac{g_{pqlmn}}{[ (p \alpha_{-2}+ q \alpha_{-1} + l \alpha_{0} + m \alpha_{1} + n \alpha_{2} )   \cdot q]^2}  .   \label{Diopot}
\end{equation}
At this point the expression is very formal, and if we wish to make this more concrete, we have the difficult task of summing up six infinite sums where the integers are restricted by the Diophantine equation (\ref{Dioph}). A more manageable version is the right hand side in (\ref{calodef}), which only involves one infinite sum and a finite sum over the representatives of the Coxeter orbits. The caveat at this point is that, while for the finite cases it is well established that these representatives can be taken to be just the simple roots with alternating signs, it is yet an open mathematical question of what the representatives are for orbits of infinite order. However, as we demonstrate now for the example of the affine Weyl group, a finite set of representatives can indeed be found. For the general $\mathbf{A}_2^{(2)}$-theory we have to construct the Coxeter element from the product of the five Weyl reflections associated to each of the simple roots: $\sigma  := \sigma_{-2} \sigma_{-1} \sigma_0 \sigma_1 \sigma_2 $. As derived in \cite{corfringinf}, closed formulae for the infinite orbits of this Coxeter element can indeed be found for the Lorentzian, hyperbolic and affine case. Here,we present just the affine case with Coxeter element $\sigma_a  := \sigma_0 \sigma_1 \sigma_2 $, for which we derived the closed formula for arbitrary powers of this element
\begin{eqnarray}
		\sigma_a^{k}(\alpha)&=& p\alpha _{-2} +q \alpha _{-1} \label{Coxeteraffgen} \\
		&& \!\!\! \!\!\!\!\!\! \!\!\!+ \left[\frac{1}{8} \left(6 k^2+1\right) q+\frac{1}{4} (6 k+3) l-\frac{1}{4} (6 k+1)
		n+\frac{m}{2}+(-1)^k\frac{2l-4m+2n-q}{8} \right] \alpha _0  \notag  \\ 
		&& \!\!\! \!\!\!\!\!\! \!\!\! +  
		\left[\frac{1}{8} \left(6 k^2-4 k-1\right) q+\frac{1}{4} (6 k+1) l-\frac{1}{4} (6 k-1)
		n+\frac{m}{2}-(-1)^k \frac{2l-4m+2n-q}{8} \right]\alpha _1  \notag\\
		&& \!\!\! \!\!\! \!\!\! \!\!\!+ \left[   \frac{1}{8} \left(6 k^2-8 k+1\right) q+\frac{1}{4} (6 k-1) l-\frac{1}{4} (6 k-3)
		n+\frac{m}{2} +(-1)^k \frac{2l-4m+2n-q}{8}  \right] \alpha _2 . \notag
\end{eqnarray} 
Next we have to select the representative of an orbit. Let us assume we can take $\alpha_2$ and consider therefore the part of the potential build on this root
\begin{equation}
	V_2(q) = \sum_{n=-\infty}^{\infty}  \frac{ g}{\left[ \sigma_a^n ( \alpha_2) \cdot q \right]^2 } . 
\end{equation}  
Using the closed formula for the orbits of the Coxeter element (\ref{Coxeteraffgen}), this part of the potential acquires the form
\begin{equation}
	V_2(q)	=\sum_{n=-\infty}^{\infty}  \frac{16 g}{\left\{2 \left[(-1)^n-1\right] q_1-2 \left[(-1)^n+1\right] q_2+\left[-6 n+(-1)^n-1\right] q_5+4 q_3\right\}^2} .
\end{equation}	
Remarkably, using the general formula $ \sum_{n=-\infty}^{\infty} (a+b n)^{-2} = \pi^2/ [\sin^2(a \pi /b) b^2]   $, this sum can be computed exactly. We obtain 
\begin{equation}
	V_2(q)	=
	\frac{\pi^2}{9 q_5^2} \left\{  \frac{g}{\sin^2\left[ \frac{\pi}{3 q_5} (q_2- q_3)  \right]   }  
	+  \frac{g}{\sin^2\left[ \frac{\pi}{3 q_5} (q_1- q_3-q_5)  \right]   }
	\right\} .
\end{equation}	
This result is encouraging as we can hope that other representative orbits can also be evaluated explicitly. Indeed using the fact that the potential should be invariant under the entire Weyl group, i.e., each individual Weyl reflection for the generators, we construct eight additional terms giving the expression
\begin{eqnarray}
		V(q) \!\!\! &=&  \!\!\! \sum_{n=-\infty}^{\infty}  \frac{ g}{\left[ \sigma^n ( \alpha_2) \cdot q \right]^2 } +
		\frac{ g}{\left[ \sigma^n ( \alpha_2) \cdot \sigma_0(q) \right]^2 }
		+ \frac{ g}{\left[ \sigma^n ( \alpha_2) \cdot \sigma_1(q) \right]^2 } \label{9terms}  \\
		&& \!\!\! + \frac{ g}{\left[ \sigma^n ( \alpha_2) \cdot \sigma_1\sigma_0(q) \right]^2 } 
		+ \frac{ g}{\left[ \sigma^n ( \alpha_2) \cdot \sigma_0\sigma_1(q) \right]^2 }
		+ \frac{ g}{\left[ \sigma^n ( \alpha_2) \cdot \sigma_0\sigma_1\sigma_0(q) \right]^2 } \notag \\
		&& \!\!\! + \frac{ g}{\left[ \sigma^n ( \alpha_2) \cdot \sigma_2\sigma_0(q) \right]^2}  
		+ \frac{ g}{\left[ \sigma^n ( \alpha_2) \cdot \sigma_2\sigma_1(q) \right]^2 }
		+ \frac{ g}{\left[ \sigma^n ( \alpha_2) \cdot \sigma_2\sigma_0\sigma_2(q) \right]^2 } \notag .
\end{eqnarray}
We have verified that the two potentials (\ref{Diopot}) and (\ref{9terms}) are indeed identical by computing explicitly the first terms in both of the expansions, i.e. to be explicit, no terms are missing and none are over counted, see \cite{corfringinf} for a list. We also stress that we did, in fact, not write down the representatives in the affine root space, but rather partly in the dual space by acting with Weyl reflections on the coordinates. Computing each term gives the potential
\begin{eqnarray}
		V(q) =  \frac{2\pi^2 g}{9 q_5^2} \left( V_{12}+ V_{13}+  V_{23} + V_{125}^+ + V_{125}^- + V_{135}^+  + V_{135}^- + V_{235}^+ +  V_{235}^-    \right) ,
\end{eqnarray}
where we abbreviated
 \begin{equation}
		V_{ij}:= \frac{1}{ \sin^2 \left[ \frac{\pi}{3 q_5} (q_i- q_j)  \right] }, \qquad
		V_{ijk}^\pm:= \frac{1}{\sin^2\left[ \frac{\pi}{3 q_5} (q_i- q_j \pm q_k)  \right]},   \qquad   i,j,k= 1,2,3,4,5 .
\end{equation}
 Next, we verify that all the terms are actually mapped into each other
 \begin{equation}
 	\begin{tabular}{ l  l l l l l }
 		$\sigma_0: $&$ V_{12} \leftrightarrow V_{235}^+  $ \,\,&$ V_{13} \leftrightarrow V_{135}^-  $\,\, &$  V_{23} \leftrightarrow V_{125}^+ $ \,\, &$  V_{125}^- \leftrightarrow V_{235}^-$ \,\,
 		& $  V_{135}^+ \leftrightarrow V_{135}^+$ , \\ 
 		$\sigma_1: $&$ V_{13} \leftrightarrow V_{23}  $ \,\,&$ V_{125}^+ \leftrightarrow V_{125}^-  $\,\, &$  V_{135}^- \leftrightarrow V_{235}^- $ \,\, &$  V_{135}^+ \leftrightarrow V_{235}^+$ \,\,
 		& $  V_{12} \leftrightarrow   V_{12} $ ,\\
 		$\sigma_2: $&$ V_{12} \leftrightarrow V_{13}  $ \,\,&$ V_{125}^+ \leftrightarrow V_{135}^+  $\,\, &$  V_{235}^+ \leftrightarrow V_{235}^- $ \,\, &$  V_{125}^- \leftrightarrow V_{135}^-$ \,\,
 		& $  V_{23} \leftrightarrow V_{23}  $  ,
 	\end{tabular}
 \end{equation}
which establishes the overall invariance of the potential under the infinite dimensional affine Weyl group. There exists an interesting limit in which we recover the standard $A_2$-Calogero model 
\begin{equation}
	\lim_{q_5 \rightarrow \pm \infty} V(q) = 2 g \left[  \frac{1}{(x_1 - x_2)^2 } +  \frac{1}{(x_1 - x_3)^2 }  + \frac{1}{(x_2 - x_3)^2 }       \right] .
\end{equation}
For the hyperbolic and Lorentzian Weyl groups we have also obtained closed algebraic formulae for the action of the respective Coxeter elements in \cite{corfringinf}, but did not yet succeed in finding all the representatives and summing up the potentials, which remain in the form of infinite sums.

\section{Conclusions and many open questions}

We have demonstrated that it is indeed possible to formulate interesting physical models based on the set of simple Lorentzian roots or even the corresponding entire infinite-dimensional root space. There are many open questions left regarding the mathematics of the infinite groups and algebras, but also concerning the physical properties of the models.

One of the more immediate mathematical issues to settle is the question of identifying the representatives for the non-intersecting orbits that produce the entire root space when acted upon with Coxeter elements of infinite order. Other mathematical questions include the still outstanding classification of the Lorentzian algebras, and of course, we are still lacking the full explicit representation, analogous to the vertex operator representation, even for simple hyperbolic algebras, let alone the Lorentzian extensions.

The proposed Lorentzian Toda field theories can be seen as a systematic framework of perturbed integrable systems, but of course, there are many more possibilities besides the option we have introduced here. Many more properties of these models can be explored, and a proper quantum version of these systems could be developed, involving the computation of the quantum corrections to the masses from renormalization, the three-point couplings, the scattering matrices, form factors, correlation functions, etc. Regarding the Calogero models, it would be very interesting to explore whether the sums in the potentials can be carried out for other algebras as well, and ultimately exploit the infinite symmetries in the construction of their solutions in the quantum theory. As a further development, one may investigate the complex ${\cal PT}$-symmetric versions of all proposed models. An interesting question is whether the complex solutions, which are now not expected to be solitons due to the lack of integrability, have also real energies. ${\cal PT}$-symmetric versions of the Calogero model may be obtained by replacing the standard roots $\alpha_i$ by their deformed versions $\tilde{\alpha}_i$ as constructed in \cite{AFZ,Mon1,fring2012non}. 
\smallskip

\noindent \textbf{Acknowledgments:} I would like to thank all my collaborators, especially Sam Whittington, Octavio Quintana, and Francisco Correa, for the material presented here.

\newif\ifabfull\abfulltrue

\end{document}